\documentclass[twocolumn]{svjour3}
\usepackage{psfrag,graphicx}
\usepackage{amsfonts,amssymb,amsmath}      
\usepackage{graphicx}
\usepackage{dcolumn}
\usepackage{bm}
\usepackage{float}
\usepackage{hyperref}
\usepackage{bbding}
\usepackage{color,soul}
\usepackage{epstopdf}

\newcommand{\erf}[1]{Eq.~(\ref{#1})}
\newcommand{\beq}{\begin{equation}}
\newcommand{\eeq}{\end{equation}}
\newcommand{\nn}{\nonumber}

\newcommand{\erfs}[2]{Eqs.~(\ref{#1})--(\ref{#2})}

\newcommand{\dg}{^\dagger}

\newcommand{\bra}[1]{\langle{#1}|}
\newcommand{\ket}[1]{|{#1}\rangle}

\newcommand{\sch}{Schr\"odinger}

\newcommand{\dbd}[1]{{\partial}/{\partial {#1}}}

\newcommand{\Tr}{\text{Tr}}

\newcommand{\tp}{^{\top}}

\renewcommand{\c}{_{\text{C}}}
\newcommand{\ob}{_{\text{o}}}
\newcommand{\un}{_{\text{u}}}

\newcommand{\ex}[1]{\langle{#1}\rangle}
\newcommand{\dd}{{\rm d}}

\newcommand{\SHUR}{\sch-Heisenberg uncertainty relation}

\newcommand{\past}[1]{\overleftarrow{#1}}
\newcommand{\fut}[1]{\overrightarrow{#1}}
\newcommand{\both}[1]{\overleftrightarrow{#1}}
\newcommand{\fil}{_{\text F}}
\newcommand{\rfil}{_{\text R}}
\newcommand{\sm}{_{\text S}}
\newcommand{\swv}{_{\rm SWV}}
\newcommand{\god}{_{\text T}}
\newcommand{\inv}{^{-1}}
\newcommand{\bx}{{\bf x}}
\newcommand{\bcx}{{\check{ \bf x}}}
\newcommand{\by}{{\bf y}}

\newcommand{\bw}{{\bf w}}

\newcommand{\xfil}{\ex{\hat\bx}\fil}

\newcommand{\xgod}{\ex{\hat \bx}\god}
\newcommand{\K}{{\cal K}}

\definecolor{nblue}{rgb}{0.06,0.3,0.73}
\definecolor{nblack}{rgb}{0,0,0}
\definecolor{nred}{rgb}{0.9,0.1,0.1}
\definecolor{nmagenta}{rgb}{0.7,0.0,0.3}

\newcommand{\blk}{\color{nblack}}

\newcommand{\hbx}{\hat\bx}

\newcommand{\Sclass}{$\mathfrak{S}$-class}

\begin{document}

\title{General criteria for quantum state smoothing with necessary and sufficient criteria for linear Gaussian quantum systems}

\author{Kiarn T. Laverick \and Areeya Chantasri \and Howard M. Wiseman}
\institute{Centre for Quantum Computation and Communication Technology 
(Australian Research Council) \and Centre for Quantum Dynamics, Griffith University, Nathan, Queensland 
4111, Australia\\ \email{kiarn.laverick@griffithuni.edu.au}\\\email{a.chantasri@griffith.edu.au} \\ \email{h.wiseman@griffith.edu.au}}

\titlerunning{Criteria for quantum state smoothing}
\authorrunning{K.~T.~Laverick et.~al.}

\date{\today}
\maketitle

\begin{abstract}
Quantum state smoothing is a technique for estimating the quantum state of a partially observed 
quantum system {\blk at time $\tau$}, conditioned on an entire observed measurement record 
(both before and after $\tau$). 
However, this smoothing technique requires an observer (Alice, say) to know the nature of
the measurement records that are unknown to her 
in order to characterize the possible true states for {\blk Bob's (say) systems}. 
If Alice makes an incorrect assumption about
the set of true states {\blk for Bob's system}, she will obtain a smoothed state that is suboptimal, 
and, worse, may 
be unrealizable (not corresponding to a valid evolution for the true states) or even unphysical (not 
represented by a state matrix $\rho\geq0$). In this paper, we review the 
historical background to quantum state smoothing, and list general criteria a smoothed quantum state 
should satisfy. Then we derive, for the case of linear Gaussian quantum systems, a necessary and 
sufficient constraint for realizability on the covariance matrix of the true state. 
Naturally, a realizable covariance of the true state guarantees a smoothed state which is physical. It might 
be thought that any putative true covariance which gives a physical smoothed state would be a realizable 
true covariance, but we show explicitly that this is not so. This underlines the importance of the realizabilty 
constraint. 
\end{abstract}

\section{Introduction}
Estimating the state of an open quantum system based on continuous-in-time measurement results is 
currently an important task in quantum science. Quantum state filtering \cite{Belav87,Bel92}, also referred 
to as quantum trajectory theory, gives the optimal estimate of the quantum state based on the 
measurement record up until the estimation time. The quantum state smoothing theory of 
Ref.~\cite{GueWis15}, on the other hand, estimates the quantum 
state based on an entire (both prior and posterior to the estimation time) measurement record. However, 
obtaining a valid smoothing theory, that is, a theory that results in a physical quantum state $\rho$, 
satisfying $\Tr[\rho] = 1$ and $\rho\geq0$ was not a trivial task 
\cite{ABL64,Tsa09a,Tsa09b,GJM13,Cha13,GueWis15}.

In order to obtain physical smoothed quantum states, Guevara and Wiseman \cite{GueWis15} 
considered a quantum system that is only partially observed by an observer. The state assigned by this 
observer, say Alice, will in general differ from the {\em true} state, i.e.,~the most accurate estimate of the 
quantum state assigned by an omniscient observer, say Bob. Bob is assumed to know two 
measurement records, one that is known to Alice (`observed' record) and 
one that is hidden to her (`unobserved' record). Even though Alice has no 
access to the unobserved record, she can still consider all possible unobserved records, and how likely 
each is, to calculate a smoothed quantum state.  

For Alice to make an optimal smoothed estimate, she must know the type of measurement that led to 
the unobserved measurement record \cite{CGW19}. However, she may not know this. 
In this work, we investigate whether it is possible for Alice to come up with some simple
physical constraints on the possible true state, that limit the set of physical smoothed states, with minimal 
assumptions. This turns out to be possible for linear Gaussian 
quantum (LGQ) systems in a steady-state regime, where the covariance matrix of the LGQ state's Wigner 
function is deterministic. Consequently, we can derive a necessary and sufficient realizabilty 
constraint \cite{WisVac01} for the set of realizable covariance matrices, assuming only that the unobserved 
measurement is fixed and diffusive in nature.

Given this new constraint on the true covariance, we investigate whether \blk it is necessary for some putative true covariance 
to satisfy this condition in order to give, according to the smoothing formula we had previously derived~\cite{LCW19}, a covariance for the smoothed state which satisfies all uncertainty relations. 
\blk We show that this is not 
the case. It is possible to select a putative true covariance that is unrealizable yet yields, 
\blk na\"ively following the procedure of Ref.~\cite{LCW19}, 
 a smoothed state that is a mathematically allowable quantum state. Indeed, it turns out that even far more stringent tests on the reasonableness of the \blk smoothed quantum state calculated using a 
putative true covariance cannot determine whether the latter satisfies the realizability 
constraint. \blk This shows the importance of the realizability constraint we have derived in 
applying the theory of quantum state smoothing. \blk  

The structure of the paper is as follows. In Sec.~\ref{Sec-His} 
we provide some historical background and some general criteria for valid quantum state smoothing 
theories. In Sec.~\ref{Sec-PRE}, we construct an argument following the Hughston-Josza-Wooters 
theorem in order to derive the necessary and sufficient realizability constraint on the true states for 
LGQ systems in steady state. Subsequently, in Sec.~\ref{Sec-PSS}, we apply increasingly tighter 
constraints on the putative true states, culminating in the newly derived realizability constraint, to 
see how they affect the physicality of the putative smoothed state derived therefrom.

\section{History and General Criteria of Quantum State Smoothing}
\label{Sec-His}
In classical estimation theory, one is typically 
tasked with assigning a value to unknown parameters of a system, 
described by a vector $\bx$, which we assume cannot be perfectly measured \cite{vanTrees1}. 
Given any information obtained from such a system, one of the most powerful tools at our disposal is a 
probability density function (PDF) $\wp(\bx)$ of the unknown, 
referred to as a {\em state} of the system. From this, one can calculate any type of estimator of $\bx$, 
such as the mean or the mode of the distribution. However, in most cases, even the state of the 
system itself can be determined in different ways, and it is first necessary to specify upon what available 
measurement data the state $\wp(\bx|C)$ is conditioned. Here `$C$' refers to any conditioning on 
measurement records. This idea underpins the field of (classical) state estimation.

For a dynamical system under continuous observation, there exist 
optimal state estimation techniques, depending on the amount of measurement information available in 
time \cite{vanTrees1,Ein12}. 
If the observed information is only available up until the time of estimation $\tau$, 
e.g.~a real-time estimation, 
we can obtain a {\em filtered} estimate $\wp\fil(\bx) := \wp(\bx|\past{\bf O})$ as a state conditioned on the 
`past' measurement record $\past{\bf O} = \{O_t;t\in[t_0,\tau)\}$. 
The adjoint problem to the filtered state is the {\em retrofiltered} effect, 
commonly referred to as the likelihood function in the literature \cite{Ein12,BroHwa12,Sarkka13},
defined as $E\rfil(\bx) := \wp(\fut{\bf O}|\bx)$, where $\fut{\bf O} = \{O_t;t\in[\tau,T)\}$. The effect tells us the 
likelihood of the `future' measurement record $\fut{\bf O}$ given the system 
parameter $\bx$ at time $\tau$. Finally, we can combine the filtered state and the retrofiltered effect 
to obtain a state conditioned on the entire, both past and future, measurement record 
$\both{\bf O} = \{O_t;t\in[t_0,T)\}$. This is 
known as the {\em smoothed} state, defined as 
$\wp\sm(\bx):= \wp(\bx|\both{\bf O}) \propto E\rfil(\bx)\wp\fil(\bx)$. Typically 
the smoothed state, when real-time estimation is not required, is more accurate 
\cite{Hay01,vanTrees1,Ein12} 
than the filtered state, as it is conditioned on more information. 

When transitioning to quantum systems, we are concerned with estimating the {\em quantum state} of the 
system as it can be considered as a quantum analogue of 
the classical PDF. For example, we can calculate a mean estimator
of any operator $\hat A$ from a quantum state $\rho$ via the expectation value 
$\ex{\hat A} = \Tr[\hat{A}\rho]$. 
By continuously monitoring the system, we can 
condition the evolution of the state on the past measurement record to obtain the filtered quantum state 
$\rho\fil$. This estimate of the state is {\blk sometimes called a} quantum trajectory 
\cite{Belav87,Bel92,WisMil10}. The quantum analog of the retrofiltered effect is 
the retrofiltered quantum effect $\hat E\rfil$, a positive operator, defined
such that $\wp(\fut{\bf O}|\rho) = \Tr[\hat{E}\rfil \rho]$ is the likelihood function for a given $\rho$. 

Following the analogy with the 
classical case, one could na\"ively combine the two quantum 
operators, the filtered state and the retrofiltered effect, to obtain a `smoothed' quantum operator,  
\beq
\varrho\sm = \frac{\hat{E}\rfil\circ\rho\fil}{\Tr[\hat{E}\rfil\circ\rho\fil]}\,, 
\eeq
where we have used the Jordan product \cite{Jordan33,JorNeuWig34}
$A\circ B = (AB + BA)/2$ to symmetrize $\varrho\sm$, and have used the denominator 
$\Tr[\hat{E}\rfil\circ\rho\fil]$ to normalise it. 
Unfortunately, the operator $\varrho\sm$ cannot properly represent a quantum state as one would hope. In 
general, this operator is not positive semidefinite, i.e.,~the criteria for a physical quantum state are not 
satisfied. Note that we are using a different notation
$\varrho$ to distinguish it from a physical quantum state denoted with $\rho$. 
Interestingly, the trace of $\varrho\sm$ with an observable $\hat A$ gives, 
\beq\label{WVE}
\Tr[\hat{A}\varrho\sm] = \frac{\Tr[\hat{E}\rfil\hat{A}\rho\fil + \rho\fil\hat{A}\hat{E}\rfil]}
{2\Tr[\hat{E}\rfil\circ\rho\fil]} = {\rm Re}[\ex{\hat A}_{\text w}]\,,
\eeq
which is not the usual expectation value of $\hat A$, but rather the real part of a complex weak value 
$\ex{\hat A}_{\rm w}$ that can have values outside the eigenvalue range of $\hat A$. The weak value was 
introduced in \cite{AAV88}, where $\rho\fil$ and $\hat{E}\rfil$ are to be interpreted as a 
generalized version 
of the pre- and post-selected states in the two-state vector formalism \cite{ABL64,Wis02,Dre10}.
It is for this reason that this $\varrho\sm$ has been referred to as the {\em smoothed weak-valued} 
(SWV) state \cite{LCW19}. 

In a similar spirit as the SWV state, another formalism, called the past quantum state (PQS) \cite{GJM13}, 
utilises the past-future measurement record in the form of the pair of operators
$\Xi({\blk \tau}) = (\rho\fil,\hat{E}\rfil)$ to estimate values of hidden results of {\blk any}
measurement performed {\blk at time $\tau$} on the quantum system. In the event that the measurement 
is weak, the estimated measurement 
reduce to the real part of the weak value, \erf{WVE} \cite{GJM13}. Since $\Xi(\tau)$ 
itself comprises two operators, it is not a direct quantum analog of the classical smoothed state.

The above raises the question: what criteria should a theory of quantum state smoothing satisfy? Here
we list four conditions:\\
\begin{itemize}
\item[(1)]{The theory should give a single smoothed quantum state $\rho\sm$ analogous to the 
classical state $\wp\sm$, and not a pair of states, for example.}
\item[(2)]{The smoothed state $\rho\sm \equiv \rho_{\both{\bf O}}$ should reduce to its corresponding 
filtered state after averaging over all 
possible future measurement records given a past measurement record, 
i.e., 
\beq
\rho\fil \equiv\rho_{\past{\bf O}} = \sum_{\fut{\bf O}}
\wp(\fut{\bf O}|\past{\bf O})\rho\sm\,.
\eeq}
\item[(3)]{\blk{The smoothed quantum state should reduce to its classical counterpart when the initial 
conditions, final conditions, and dynamics of the system can all be described probabilistically in a fixed 
basis.}}
\item[(4)]{The smoothed quantum state must be an \Sclass~quantum state; that is, it must be Hermitian and positive semidefinite.}
\end{itemize}

{\blk Formally, for a Hilbert space $\mathbb{H}$, the \Sclass~quantum states are the set  
$\mathfrak{S}(\mathbb{H}) = \{\rho\in \mathfrak{B}(\mathbb{H}): \rho\geq 0,\, \rho = \rho\dg, \Tr[\rho] = 1\}$, where $\mathfrak{B}(\mathbb{H})$ is the set of bounded linear operators on $\mathbb{H}$~\cite{Holevo}.}
We show, in Appendix~\ref{App-A}, that the SWV state satisfies 
the properties (1)--(3), but not (4). 
Moreover, it is worth noting that the properties (2) and (4)
imply that, if $\rho\fil$ is pure, then we have $\rho\sm = \rho\fil$. 
In other words, in order to obtain a non-trivial 
smoothed state, there must be missing information about the system.
With these properties in mind, Guevara and Wiseman \cite{GueWis15} devised a theory for quantum state 
smoothing, yielding a valid smoothed quantum state which satisfies all of the above propeties.

\begin{figure}[t!]
\includegraphics[scale = 0.335]{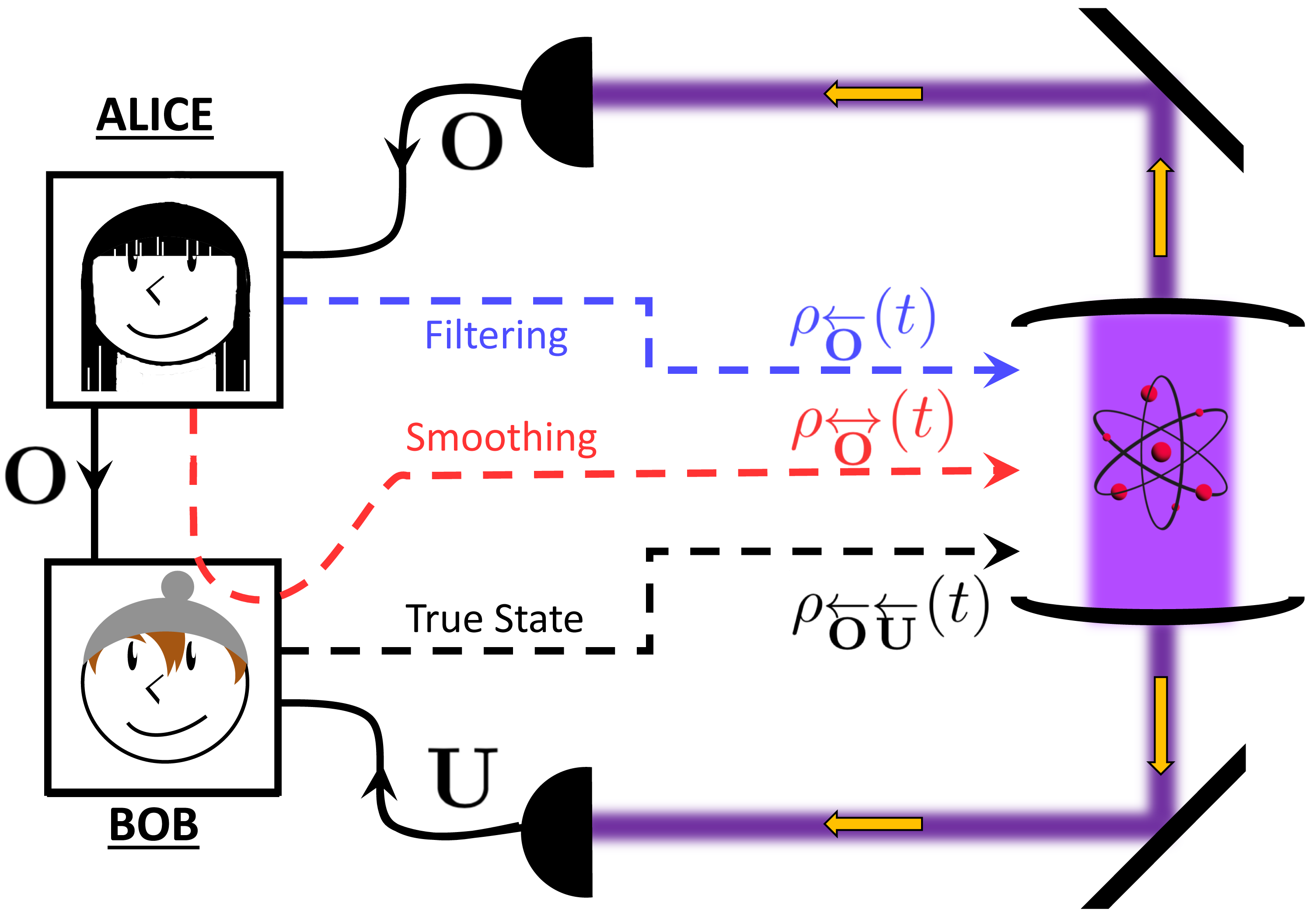}
\caption{A diagrammatic representation of the quantum state smoothing 
formalism. If Alice, who only has access to the observed record ${\bf O}$, wanted to assign a quantum 
state to the system, she could condition her estimate of the state on the past observed measurement 
record resulting in the filtered quantum state $\rho\fil:=\rho_{\protect\past{\bf O}}$. Another observer, Bob, 
can also assign a quantum state to the system conditioned on the his measurement record ${\bf U}$, in 
addition to the observed record, giving the true quantum state 
$\rho\god :=\rho_{\protect\past{\bf O},\protect\past{\bf U}}$. At this point, if Alice knows that there is an 
unobserved measurement record, she can obtain a more accurate estimate of the true quantum state of 
the system by conditioning her estimate on the past-future observed measurement record to obtain the 
smoothed quantum state $\rho\sm := \rho_{\protect\both{\bf O}}$.}
\label{Fig-QSS}
\end{figure}

The quantum state smoothing theory \cite{GueWis15} 
is defined for an open quantum system, using a scenario where the 
system is imperfectly monitored by an observer. As described in Fig.~\ref{Fig-QSS}, the observer named 
Alice observes only her measurement record $\bf{O}$, whereas an omniscient observer, Bob, has 
access both 
to Alice's record and to the information Alice missed in her measurement, {\blk which comprises} a 
measurement record $\bf{U}$ (that is unobserved by Alice). 
If Alice had access to the unobserved record, or 
equivalently made a perfect measurement on the system, she would have maximum knowledge about the 
quantum state and her estimated state would be the `true' state 
$\rho\god \equiv \rho_{\past{\bf O}\past{\bf U}}$ which Bob has. Since Alice does not 
have access to the unobserved record, her task is to best estimate Bob's state using solely her 
observed record. Alice can define her estimated state as a conditioned state,
\beq\label{cond_state}
\rho\c = \sum_{\past{\bf{U}}}\wp\c(\past{\bf{U}})\rho_{\past{\bf O}\past{\bf{U}}}\,,
\eeq
where the summation is over all possible past unobserved records $\past{\bf U}$ and 
the conditioning `$\text{C}$' can be any part of the observed record. For the case of filtering, 
$\rho\c = \rho\fil$, the conditional probability distribution of the unobserved record becomes 
$\wp\fil(\past{\bf U}) = \wp(\past{\bf U}|\past{\bf O})$. In the same spirit, a proper smoothed quantum state 
is defined \cite{GueWis15} by using the probability distribution of the unobserved record conditioned on the 
past-future measurement record, i.e., 
$\wp\sm(\past{\bf U}) = \wp(\past{\bf U}|\both{\bf O})$. As alluded to earlier 
(see Appendix~\ref{App-A}), this smoothed quantum state 
satisfies all of the properties (1)--(4) required for a quantum state smoothing theory, 
providing the first direct quantum analog of the classical smoothed state.

\section{Constraints on the underlying true states}\label{Sec-PRE}

\subsection{\blk General considerations}

While the quantum state smoothing theory satisfies all of the criteria mentioned earlier, it requires the 
unobserved unravelling performed by the secondary observer, Bob. 
Since Bob's record is unknown to Alice, it might seem like Bob's true state $\rho_{\past{\bf O},\past{\bf U}}$ 
could be any arbitrary pure 
states of the system that still satisfies \erf{cond_state}, for some $\wp_{\blk \both{\bf O}}(\past{\bf U}$). 
This idea is similar to the fact that a mixed state can be written as a combination of 
pure states in infinitely many ways. However, as {\blk mooted in the Introduction}, this is not the case. 
There are some physical constraints on Bob's possible true states to reflect the fact that the true 
state, in principle, is computed using an actual unobserved measurement record. 

Let us restrict the discussion to a steady state of a monitored quantum 
system, where the system has evolved for 
a sufficiently long time and its dynamics are independent of any 
initial condition. In this case, a true state of the system will 
certainly be a pure state. The unconditional 
steady state is 
an ensemble average of true states $\rho\god := \rho_{\past{\bf O}\past{\bf U}}$ over both the 
past observed and unobserved measurement records,
\beq\label{uncond}
\rho^{\rm ss} = \sum_{\past{\bf O},\past{\bf U}}\wp(\past{\bf O},\past{\bf U})\rho\god\,.
\eeq
Alice can condition her estimate of the quantum state on her past measurement record
$\past{\bf O}$ to obtain a filtered state
\beq\label{rhoF}
\rho\fil = \sum_{\past{\bf U}}\wp\fil(\past{\bf U})\rho\god\,,
\eeq
{\blk where $\wp\fil(\past{\bf U}) = \wp(\past{\bf U}|\past{\bf O})$ as before.}
These two mixed states, both the unconditioned and filtered states in \erfs{uncond}{rhoF}, 
can be used to restrict the allowed true states by only considering the pure states that satisfies these 
equations. However, these constraints, even though necessary, are not sufficient for the 
realizable true states, as we now show. 

The Hughston-Jozsa-Wooters (HJW) theorem provides a method for deriving an additional 
constraint. Consider a mixed quantum state of a steady state system, defined as
\beq\label{HJW-SS}
\rho^{\rm ss} = \sum_k \wp_k{\rm \hat\Pi}_k\,,
\eeq
a mixture of pure states ${\rm \hat\Pi}_k = \ket{\psi_k}\bra{\psi_k}$ with probability 
weights $\wp_k > 0$. In general, the pure states ${\rm \hat\Pi}_k$ do not need to be orthogonal, and 
there are infinitely many ensembles $\{\wp_k,{\rm \hat\Pi}_k\}_k$ that can represent the 
mixed state $\rho^{\rm ss}$. 
However, the HJW theorem states that, if a state is mixed solely because of its  
entanglement with an environment, it is possible to measure the 
environment in such a way that the system state collapses into a particular pure state ${\rm \hat \Pi}_k$ in the 
ensemble with the corresponding probability $\wp_k$, without, on average, disturbing the system. Note, 
the measurement does not have to be a projective measurement, but can, for example, be a measurement 
at any time $t$ during a continuous monitoring of the system.

We follow Ref.~\cite{WisVac01} to utilise the HJW theorem for a continuously probed system 
in steady state. Given a system in a mixed unconditioned 
state $\rho^{\rm ss}$, by measuring the environment at 
time $t$, the system's state collapses into one of the pure states 
${\rm \hat\Pi}_k$ of the ensemble $\{\wp_k,{\rm \hat\Pi}_k\}_k$. After the measurement, the system re-entangles with the 
environment for some time $\Delta t$, resulting in its state evolving into a mixed state $\rho(t+\Delta t)$ 
before the next measurement at time $t+\Delta t$. However, if 
we are to keep the same representation for the system, the measurement at time $t+\Delta t$ should 
collapse the system state to a pure state in the same ensemble. 
This is the condition for physically realizable states for continuous measurements, taking $\Delta t$ to be
an infinitesimal time difference $\dd t$.

In this work, we will adapt the physically realisability constraint derived in \cite{WisVac01} to the Alice-Bob 
protocol, replacing the unconditioned state $\rho^{\rm ss}$ with Alice's filtered state, and the ensemble of 
pure states with the ensemble of Bob's possible true states. 
At some time $t$, Alice, with her observed record 
up to time $t$, assigns her best estimate of the state to be a mixed state $\rho_{\past{\bf O}}(t)$. 
Bob, who can measure the environment unaccessible to Alice, 
collapses the system's state yielding a 
particular pure state in the ensemble of possible true states, given by \erf{rhoF},
$\{\wp\fil(\past{\bf U}),\rho\god\}_{\past{\bf U}}$. Evolving this state for some time $\dd t$
only conditioning on Alice's measurement record, 
we obtain a new mixed state. Bob then, at time $t+\dd t$, use his unobserved record for that 
time to collapse the state, again, into a possibly new true state.

At this point one might be tempted to just apply the analogous necessary condition derived in 
\cite{WisVac01} by replacing the 
unconditioned state with the filtered state and the pure state with the true state, and be done with the 
problem. However, due to the stochastic nature of the filtered state as it is conditioned on the observed 
record, it {\blk is not a fixed state}. As a 
consequence, we cannot, in general, claim that the system's mixed state at different time should
have the same representation of true states.
Nevertheless, a general constraint on the ensemble of physically realizable true states can derived for 
specific systems that have some level of determinism, where some 
properties of the filtered state remain unchanged throughout the evolution from time $t\to t+\dd t$.
An example of this is the class of LGQ systems \cite{WisMil10,DohJac99,WisDoh05} to which we will restrict our discussion to 
henceforth.

\subsection{LGQ systems}\label{Sec-LGQ}
LGQ systems are continuous-variable quantum systems that can be described by $N$ bosonic modes, or 
equivalently a $2N$ vector $\hbx = (\hat{q}_1,\hat{p}_1, ... , \hat{q}_N,\hat{p}_N)\tp$, where $\hat{q}_k$ 
and $\hat{p}_k$ are the usual position and momentum operators satisfying 
$[\hat{q}_k,\hat{p}_\ell] = i\hbar\delta_{k\ell}$.
As the name suggests, these 
systems have linear dynamics {\blk (in the sense defined below)}, and the Wigner representation of the 
quantum state is Gaussian, i.e., $W(\bcx) = g(\bcx; \ex{\hbx},V)$ with 
mean $\ex{\hbx}$ and covariance $V$. The linear{\blk ity constraint requires that the} 
unconditioned Wigner function satisfies the Ornstein-Uhlenbeck equation \cite{WisMil10}
\beq\label{OUeq}
\dot{W}(\bcx) = (-\nabla\tp A\bcx +\frac{1}{2}\nabla\tp D\nabla)W(\bcx)\,,
\eeq
where $\nabla = (\dbd{x_1},...,\dbd{x_{2N}})\tp$ with $A$ and $D$ being constant matrices. 
Furthermore, {\blk it requires that} any record $\by_{\rm r}$ resulting from a 
measurement of this LGQ system must be linear in $\hbx$; that is \cite{WisMil10,DohJac99,WisDoh05}
\beq
\by_{\rm r}\dd t = C_{\rm r}\xgod\dd t +\dd\bw_{\rm r}\,,
\eeq
where $\xgod$ is the true mean of the system, $C_{\rm r}$ is a constant matrix, 
$\rm r \in\{{\rm o},{\rm u}\}$, with `${\rm o}$' and `${\rm u}$' 
representing the observed and unobserved measurement records, respectively. Here, we have also 
introduced the measurement noise $\dd\bw_{\rm r}$, a Weiner increment satisfying 
\beq
\mathbb{E}[\dd\bw_{\rm r}] = 0\,, \qquad \dd\bw_{\rm r}\dd\bw_{\rm r'}\tp = \delta_{\rm rr'} I\dd t\,,
\eeq
where $\mathbb{E}[...]$ denotes an ensemble average {\blk and $I$ denotes the identity matrix}. By satisfying these conditions above, it is 
guaranteed that the unconditioned and conditioned states remain Gaussian throughout their evolution.

To find the set of physically realizable true states, we begin with Alice. At time $t$, Alice computes and 
assigns her (mixed) filtered state, characterised by a filtered Wigner function 
$W\fil(\bcx) = g(\bcx; \xfil,V\fil)$, to the system. The filtered mean $\xfil$ and covariance $V\fil$, conditioned 
on the past observed record $\by\ob$, are given by \cite{WisMil10,LCW19}
\begin{align}
&\dd \xfil = A\xfil \dd t + \K\ob^+[V\fil]\dd\bw\fil \,,\label{dXF} \\
&\dot{V}\fil = AV\fil + V\fil A\tp + D - \K\ob^+[V\fil]\K\ob^+[V\fil]\tp \label{dVF}\,.
\end{align}
Here $\K^{\pm}\ob[V] = VC\ob\tp \pm\Gamma\ob\tp$ and we have introduced the matrix $\Gamma\ob$ to 
account for the measurement back-action from the observed measurement record $\by\ob$. 
The stochastic nature of the mean can be seen in \erf{dXF} 
through the vector of innovations $\dd\bw\fil = \by\ob\dd t - C\ob\xfil\dd t$, which is a stochastic quantity 
describing the difference between the measurement result and the calculated estimate. On the 
other hand, the filtered covariance \erf{dVF} is entirely deterministic and is fixed by the choice of Alice's 
measurement unravelling \cite{WisMil10,DohJac99,WisDoh05}.

Since we are interested in the steady state, we assume that the eigenvalues of the drift matrix $A$ in 
\erf{OUeq} are negative and the time $t$ is sufficiently large that the system has reached its steady state. 
We can set the left-hand side of \erf{dVF} to zero to 
solve for the steady-state solution $V\fil^{\rm ss}$. With Bob making the measurement unobserved by 
Alice, the state collapses into a particular pure 
state $W\god(\bcx) = g(\bcx;\xgod,V^{\rm ss}\god)$ with mean $\xgod$ and covariance $V\god^{\rm ss}$. 
This true state will have a purity of unity, 
where purity for a Gaussian state is defined as $P = (\hbar/2)^N\sqrt{{\rm det}(V\inv)}$ for a covariance 
$V$ \cite{WisMil10}. 
To say that this state is one of the pure states in the mixture of Alice's filtered state, we require that 
\beq\label{FiltCond}
V^{\rm ss}\fil - V^{\rm ss}\god \geq 0\,,
\eeq
meaning that $V^{\rm ss}\god$ fits within $V^{\rm ss}\fil$ in $2N$-dimensional phase space.

Following the scheme presented in the previous subsection for the physically realizable states, the state 
$W\god(\bcx)$ evolves with \erfs{dXF}{dVF} for a time $\dd t$ to become a {\blk filtered} state. 
The updated mean and variance of the new {\blk filtered} state is given by {\blk
\beq
\begin{split}\label{evolved-XT}
\xfil'(t+\dd t) = \xgod(t) + A&\xgod(t)\dd t + \\&\K\ob^+[V\fil'(t)]\dd\bw\ob(t)\,,
\end{split}
\eeq
\beq
\begin{split}\label{evolved-VT}
V\fil'(t+\dd t) = V\fil'(t) + (A&V\fil'(t) + V\fil'(t) A\tp + D - \\&
\K^+\ob[V\fil'(t)]\K^+\ob[V\fil'(t)]\tp)\dd t\,,
\end{split}
\eeq
where $V\fil'(t) = V^{\rm ss}\god$ and the prime is to distinguish this filtered mean and covariance from that 
in \erfs{dXF}{dVF}. Since this state at time $t+ \dd t$ is a filtered state, by definition, it is a mixture of true 
states at $t$. As a result, it must be the case that $V\fil'(t+\dd t) - V^{\rm ss}\god \geq 0$, and from 
\erf{evolved-VT} we obtain }
\beq\label{True_PR}
AV^{\rm ss}\god + V^{\rm ss}\god A\tp + D - \K^+\ob[V^{\rm ss}\god]\K^+\ob[V^{\rm ss}\god]\tp\geq 0\,.
\eeq
This condition gives the set of realizable true covariances, which is the main result of this paper.

We can also show that \erf{True_PR} is a sufficient constraint for the true covariance by showing that a 
pure covariance that satisfies \erf{True_PR} must be the steady-state true covariance $V\god^{\rm ss}$. 
Beginning with a pure true state with a Wigner function 
$W\god(\bcx) = g(\bcx,\xgod,V\god)$ such that $V\god$ that satisfies 
\erf{True_PR} at time $t$, the true state at an infinitesimal time later, $t+\dd t$, when only conditioning on 
Alice's observed record, must evolve into a mixture of pure states, each with covariance $V\god$ 
and Gaussian-distributed means. We know, by the HJW theorem, that Bob can measure the 
system in some way to collapse the system into one of the pure states in the ensemble with the 
appropriate probability and the resulting covariance of the collapsed pure state at time $t+\dd t$ will be 
$V\god$. Thus, Bob can always re-prepare a pure state with covariance $V\god$ if he 
continuously monitors the system over the time interval $[t,t+\dd t)$ and, consequently, since the 
covariance remains unchanged over the time interval, $V\god$ must be the steady state solution 
$V^{\rm ss}\god$ by definition. This means that \erf{True_PR} is necessary and sufficient for the true 
covariance in LGQ systems.

\begin{figure*}[t!]
\begin{minipage}{.5\textwidth}
\includegraphics[scale = 0.32]{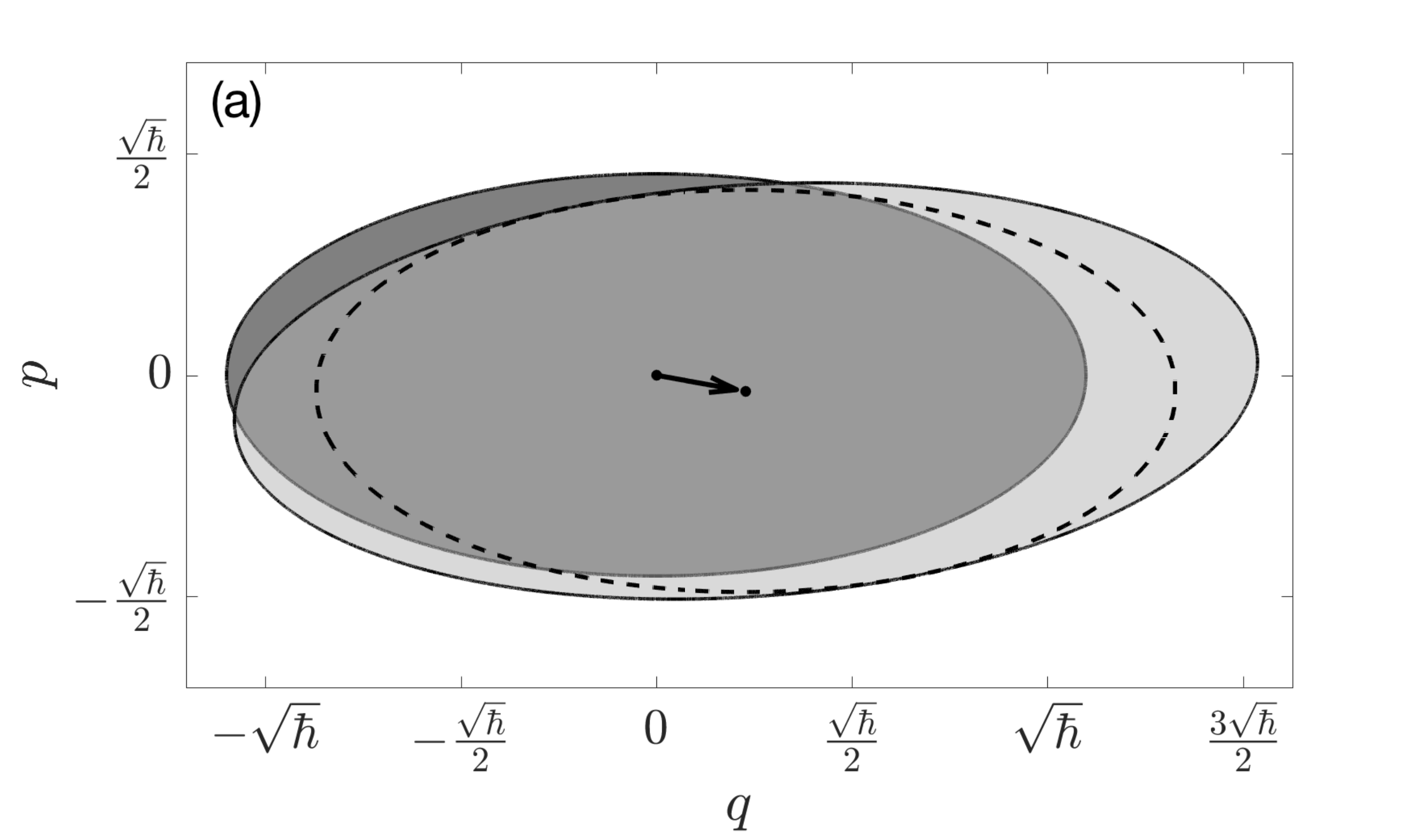}
\includegraphics[scale = 0.32]{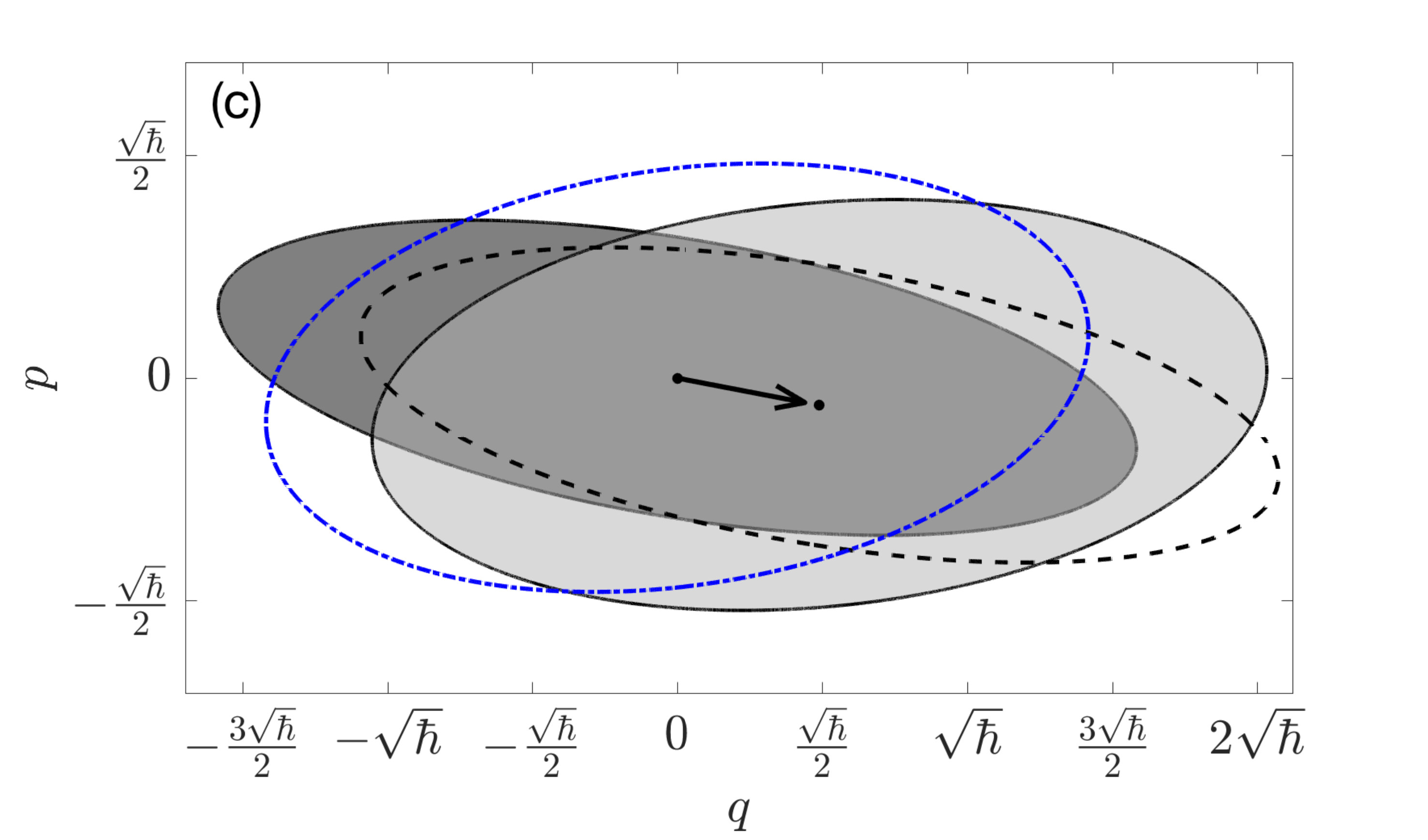}
\end{minipage}%
\begin{minipage}{.5\textwidth}
\includegraphics[scale = 0.32]{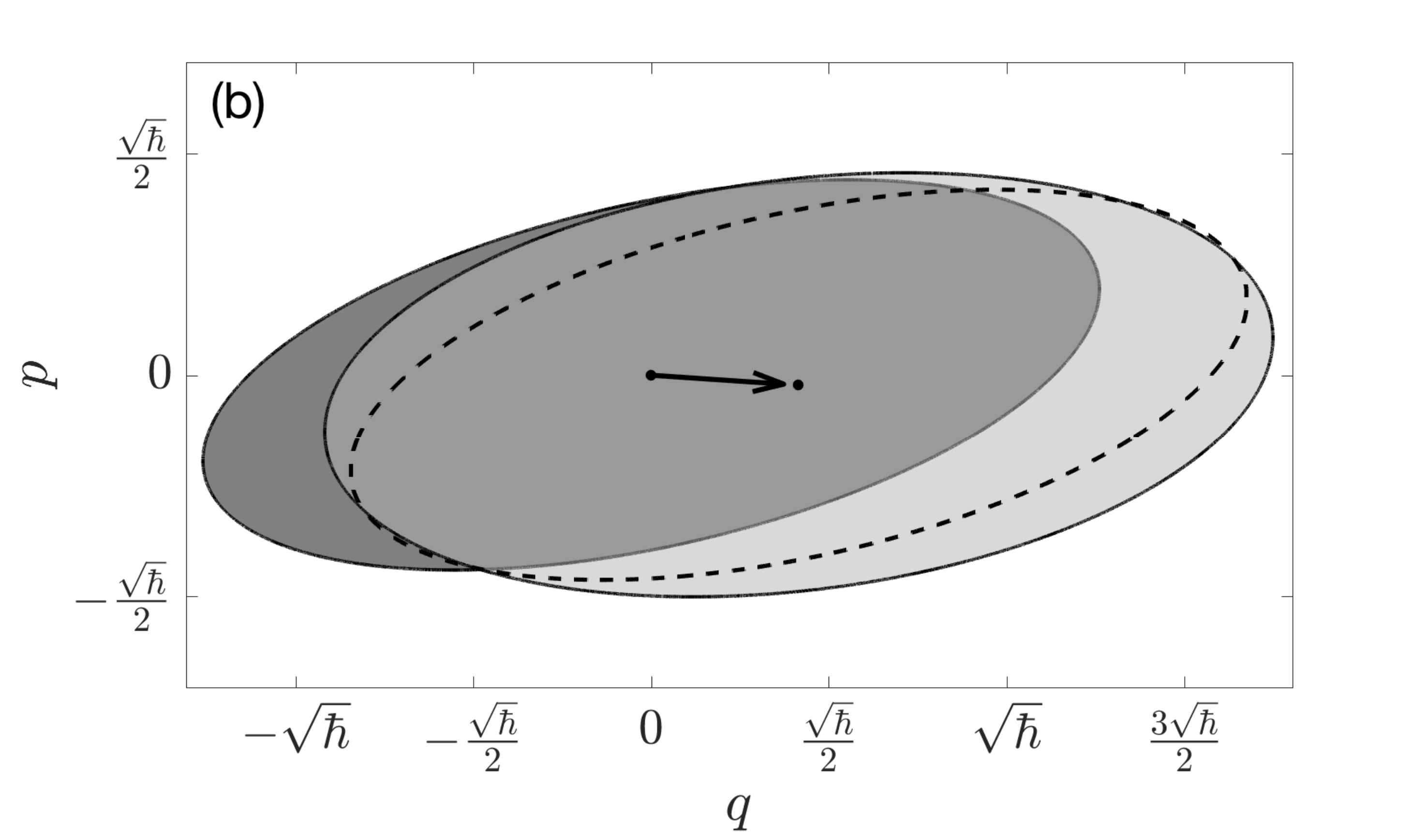}
\includegraphics[scale = 0.32]{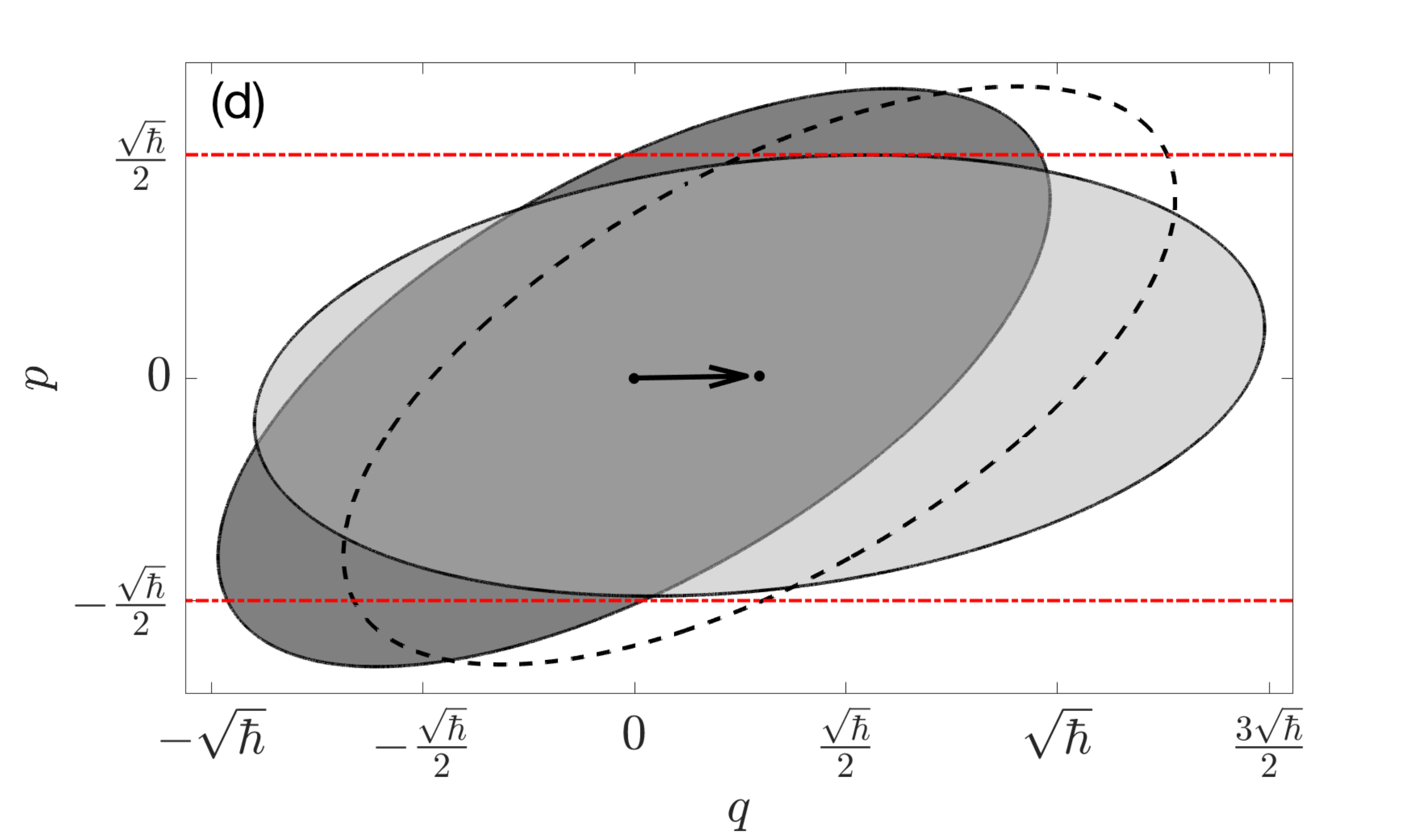}
\end{minipage}
\caption{The phase space diagrams for 1-SD contours of the Wigner functions for the OPO systems, 
showing the physical realisability condition for the Alice-Bob scenario. \blk (See text for parameter details.) \blk  In all four graphs, (a)--(d), four cases 
are plotted with different $V^{\rm ss}\god = V^{(i)}$ for $i\in\{a,b,c,d\}$, 
shown in \erf{test-V1}. The {\blk putative} covariances $V^{\rm ss}\god$ at 
time $t_0 = 0$ are plotted as the dark-grey ellipses, and are replotted as dashed ellipses with the means 
translated to their new values after evolving for a time $\Delta t = 0.8$ s, 
in order to compare with the evolved covariances (light-grey ellipses) at 
the new time $t_0 + \Delta t$ from \erf{evolved-VT}. The mean values evolve according to \erf{evolved-XT}.
Only one case, (a), shows that the initial covariance fits inside its evolved state, because 
it satisfies \erf{True_PR}. In the remaining cases, the covariances at $t_0 + \Delta t$ have some regions 
outside their evolved states. For the cases (c) and (d), the initial covariances do not even fit within the 
filtered (blue dot-dashed) and the unconditioned (red dot-dashed) steady state covariances, respectively.}
\label{Fig-PSD}
\end{figure*}

\subsection{\blk Example: optical parametric oscillator}\label{ss:ex}

We will now illustrate the effect of this constraint, \erf{True_PR}, by considering a physical model, the 
on-threshold optical parametric oscillator (OPO) \cite{WisMil10,WisDoh05}. The OPO system is described 
by the Lindblad master equation 
\beq\label{OPO}
\hbar\dot{\rho} = -i[\hat{q}\hat{p}+\hat{p}\hat{q},\rho] + {\cal D}[\hat{q}+i\hat{p}]\rho\,,
\eeq
where ${\cal D}[\hat{c}]\bullet = \hat{c}\bullet\hat{c}\dg -\{\hat{c}\dg\hat{c}/2,\bullet\}$. This master equation 
yields $A = {\rm diag}(0,-2)$ and $D = \hbar I$, where $I$ is the $2\times2$ identity matrix. 
We also consider a homodyne measurement for Alice, giving 
$C\ob = 2\sqrt{\eta\ob/\hbar} (\cos\theta\ob, \sin\theta\ob)$ and $\Gamma\ob = -\hbar C\ob/2$, 
where $\theta\ob$ and 
$\eta\ob$ are the homodyne phase of Alice's measurement and its efficiency. 
For this specific example we will consider $\theta\ob = 3\pi/8$ 
and $\eta\ob = 0.5$.

We consider four putative covariances for Bob, $V^{\rm ss}\god = V^{(i)}$ 
for $i\in\{a,b,c,d\}$. The specific covariance matrices are 
\beq\label{test-V1}
V^{(a)} = \frac{\hbar}{2}\left[\begin{array}{cc}
2.41 & 0\\
0 & 0.41
\end{array}\right]\,,\,\, V^{(b)} = \frac{\hbar}{2}\left[\begin{array}{cc}
3.18 & 0.49\\
0.49 & 0.39
\end{array}\right]\,,
\eeq
\beq
V^{(c)} = \frac{\hbar}{3}\left[\begin{array}{cc}
5.02 & -0.50\\
-0.50. & 0.25
\end{array}\right]\,,\,\, V^{(d)} = \frac{\hbar}{2}\left[\begin{array}{cc}
1.93 & 0.79\\
0.79 & 0.84
\end{array}\right]\,,\nn \label{sample-V}
\eeq
where the only covariance that satisfies the realizablility constraint \erf{True_PR} is $V^{(a)}$. 
Covariance $V^{(b)}$ was chosen to satisfy \erf{FiltCond} and $V^{(c)}$ was chosen to satisfy
\beq\label{UncCond}
V^{\rm ss} - V^{\rm ss}\god \geq 0\,,
\eeq 
but not \erf{FiltCond}, which will be useful for the discussion in the next section. 
Finally, $V^{(d)}$ does not satisfy \erf{UncCond}, but is still a pure state. In Fig.~\ref{Fig-PSD}, we 
observe how these four states, with an initial mean $\xgod = (0,0)\tp$, evolve under 
\erfs{evolved-XT}{evolved-VT}. The only covariance that still fits within the evolved state is $V^{(a)}$, as 
expected since this satisfies \erf{True_PR}. The remaining cases all have some region where the initial 
covariance is outside the evolved covariance (more detail is in the caption of Fig.~\ref{Fig-PSD}).

\section{What the smoothed state reveals about physical realizability}
\label{Sec-PSS}
Now that we know the necessary and sufficient constraint for the true covariance to be physically realizable, 
the reader might be 
wondering whether a violation of this constraint can be detected directly from the smoothed quantum state. 
Explicitly, we ask the question: \blk if Alice were told a {\em putative} true covariance $\tilde{V}\god$ by Bob, and she used that, innocently, to calculate her smoothed state following the formulae in Ref.~\cite{LCW19}, 
would she be able to tell from the nature of that smoothed state whether the $\tilde{V}\god$ she had been told could or could not be the actual true covariance. \blk

Consider the same example as in the preceding section, Sec.~\ref{ss:ex},  the OPO system in \erf{OPO}, with the same parameters for the observed record as before ($\eta\ob = 1/2$ and  $\theta\ob = 3\pi/8$), to help guide the analysis in this section. 
We can represent a putative true covariance as a $2\times2$ matrix,
\beq\label{pVT}
\tilde{V}\god = \frac{\hbar}{2}\left[\begin{array}{cc}
\alpha & \beta\\
\beta & \gamma
\end{array}\right]\,.
\eeq
\blk Since the true state is pure, this will satisfy $\alpha\gamma - \beta^2 = 1$, so its parameters can be reduced to two, say $\gamma$ and $\delta = \beta/\sqrt{\alpha\gamma}$ \cite{WisMil10}. 

\begin{figure}
\includegraphics[scale=0.38]{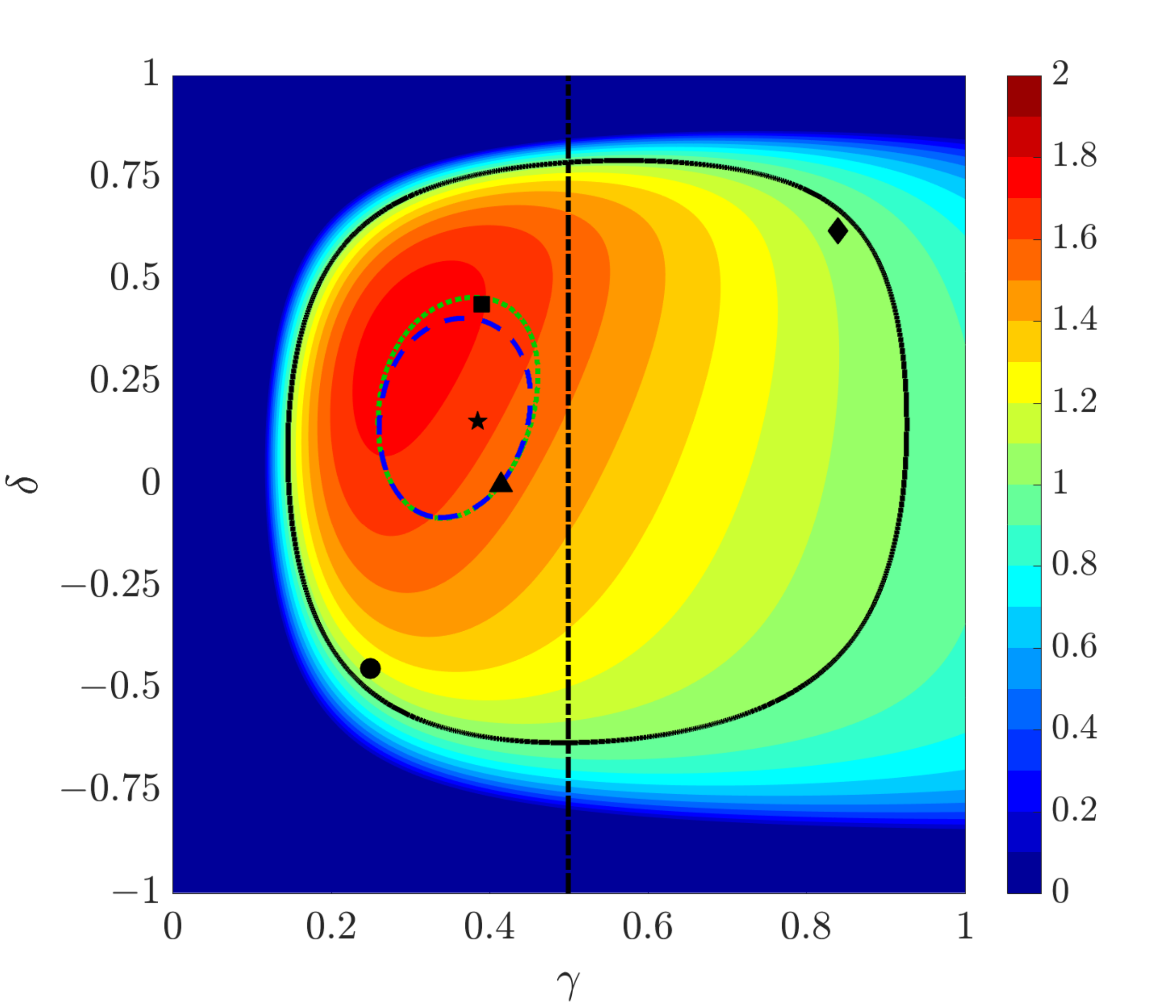}
\caption{\blk A portion of the set of putative true covariances $\tilde{V}\god$ in \erf{pVT}, where we have defined $\delta = \beta/\sqrt{\alpha\gamma}$, for the example OPO system. (See text for parameter details.) The dot-dashed line separates the 
covariances that satisfy \erf{UncCond} (area on the left). The green
dotted and blue dashed lines indicates the covariances satisfying \erf{FiltCond} and \erf{True_PR}, 
respectively. The colours indicate the value of $(2/\hbar)^2{\rm det}(V^{\rm ss}\sm)$ for the smoothed state covariance $V^{\rm ss}\sm$ calculated from $\tilde{V}\god$ using the theory of Ref.~\cite{LCW19}. Values greater than one  ($\mathfrak{S}$-class smoothed states) obtain inside the solid black line. \blk The triangle, square, circle and diamond markers indicate the covariances matrices $V^{(a)}$, $V^{(b)}$, $V^{(c)}$ and 
$V^{(d)}$ in \erf{test-V1}, respectively. We have also considered two particular unobserved unravellings 
for the system. The first is a homodyne unravelling with a phase $\theta\un = -\pi/8$, indicated by the 
triangle marker, and a balanced heterodyne unravelling indicated by the star marker.}
\label{Fig-trucond}
\end{figure}

In Fig.~\ref{Fig-trucond} we plot, using the parameters $\gamma$ and $\delta$, a region in the parameter space of putative true state covariances, and the three conditions on it considered in Sec.~\ref{Sec-LGQ}. The weakest necessary condition is \erf{UncCond}, meaning that the true state fits, in $2N$-dimensional phase space, within the unconditioned steady state. Here this corresponds to the constraint $\gamma < 0.5$ 
(left of the dash-dot line). Within that region is the region (bounded by the dotted green line) of 
putative true states satisfying the stronger necessary condition is \erf{FiltCond}, meaning that the true state fits within the steady-state filtered state. Within that region is the region (bounded by the dashed blue line) of putative true states satisfying the strongest necessary condition, also a sufficient condition, \erf{True_PR}. 

Now for quantum state smoothing, Alice calculates the covariance of her smoothed state, in steady state, directly from the putative true covariance via \cite{LCW19} \blk
\beq\label{VS}
V^{\rm ss}\sm = [(V^{\rm ss}\fil-V^{\rm ss}\god)\inv + (V^{\rm ss}\rfil+V^{\rm ss}\god)\inv]\inv +
V^{\rm ss}\god\,,
\eeq
where the retrofiltered covariance $V^{\rm ss}\rfil$ is the solution to the adjoint equation of \erf{dVF}, 
i.e.,~in the steady state,
\beq
AV^{\rm ss}\rfil + V^{\rm ss}\rfil A\tp + D - \K\ob^-[V^{\rm ss}\rfil]\K\ob^-[V^{\rm ss}\rfil]\tp = 0\,.
\eeq
\blk  
The question is: does $V^{\rm ss}\sm$ straight-fowwardly reveal whether $\tilde{V}\god$ (and hence $V^{\rm ss}\sm$) is physically realizable? 

To begin, we can look at when $V^{\rm ss}\sm$ is an $\mathfrak{S}$-class state. For \blk Gaussian \blk systems, the necessary and sufficient criteria for an 
\Sclass~state 
 is that the covariance matrix $V$ satisfies the \SHUR\ $V + i\hbar\Sigma/2 \geq 0$ 
\cite{WisMil10,Holevo,Bra05}, where $\Sigma = \oplus^N\left[\begin{smallmatrix}
0 & 1\\
-1 & 0
\end{smallmatrix}\right]$. For $N=1$ (as here) this reduces to ${\rm det}(V) \geq \hbar^2/4$. In Fig.~\ref{Fig-trucond} we plot  ${\rm det}(V^{\rm ss}\sm)/(\hbar^2/4)$ for $V^{\rm ss}\sm$ as a function of the putative true state covariance parameters from \erf{VS}. The $\tilde{V}\god$ that give rise to $\mathfrak{S}$-class 
smoothed states are inside the solid black line. Clearly, the set of 
putative true covariances that result in $\mathfrak{S}$-class smoothed states is {\em not} restricted to the 
realizable true covariances as \blk a good fraction (the area on the right side of the dot-dashed line) \blk of these putative true covariances do not even satisfy \erf{UncCond}. (Note however that here \erf{FiltCond} is sufficient for the generated smoothed state to be \Sclass; see Appendix~\ref{App-B} for the proof that this is the case for all LGQ systems.) \blk 

One might wonder whether putting further restrictions on $V^{\rm ss}\sm$ would help the situation. Since it is, by definition (\ref{cond_state}), a mixture of true states, $V^{\rm ss}\sm$ should also fit inside both the unconditioned state $V^{\rm ss}$ and the filtered state $V^{\rm ss}\fil$. That is \erf{UncCond} and \erf{FiltCond} should both be satisfied with $V^{\rm ss}\sm$ in place of $\tilde{V}\god$. Surprisingly, imposing these extra constraints makes no difference. That is, $V^{\rm ss}\sm$ by construction is guaranteed to satisfy these constraints even when $\tilde{V}\god$ does not. This is proven in Appendix~\ref{App-C} for arbitrary LGQ systems. This is so because the equation for calculating the smoothed quantum state in Ref.~\cite{LCW19} is oblivious to the unphysicality of the matrices which appear in it. Matrices which should be positive semidefinite, like $V^{\rm ss}\fil - \tilde V^{\rm ss}\god$, may become indefinite without making the smoothed state obviously wrong. That is, Alice should not innocently accept a putative true covariance $\tilde V^{\rm ss}\god$ told to her by Bob, but should first check whether it satisfies the necessary and sufficient condition \erf{True_PR}. \blk 


\blk In Fig.~\ref{Fig-trucond} we have also pointed out the four example putative true states considered in Sec.~\ref{ss:ex}. As this shows, all four give rise to \Sclass~smoothed states, even though only one satisfies \erf{True_PR}. This, indicated by the triangle marker in 
Fig.~\ref{Fig-trucond}, corresponds to $V^{(a)}$ from \erf{test-V1}. It results from a 
homodyne measurement by Bob with phase $\theta\un = -\pi/8$, and is an extremal point of the set of physically realizable $V\god$. In 
fact the set of $V\god$ generated by homodyne measurements for Bob, for all possible phases, forms the set 
of extremal points of the physically realizable set. \blk (Note that this is not the case in general; it is  
a property 
of systems with a single mode ($N = 1$) and a single Lindblad operator $\hat{c}$). To illustrate this 
we have also considered a balanced heterodyne measurement \cite{WisMil10} by Bob, \blk indicated by the star 
marker, which is in the interior of the set of realizable covariances.

\section{Conclusion}
In this paper we reviewed the history that lead to the development of the quantum state smoothing theory  
\cite{GueWis15}, beginning from the classical notions of smoothing, and the 
motivation behind the theory. In formulating the smoothed quantum state, a true quantum state had to be 
introduced, which depends on a hidden measurement record. This motivated the 
investigation of the necessary constraints on the true state of the system when an observer has 
access to only the observed measurement record. {\blk A simple} constraint is impossible {\blk to 
formulate} general, as the evolution of the quantum system under observation is inherently stochastic and 
never reaches a steady state. However, for specific systems, like the LGQ systems we 
considered, there is enough determinism in the system to allow us to derive a necessary and sufficient 
constraint on the physically realizable true states. 
Finally, we find that the {\blk mathematical validity of a smoothed state calculated 
na\"ively} from a putative true 
covariance can not witness whether that covariance satisfied the physical realisability constraint. 
{\blk Nor do further, more stringent, conditions on the smoothed state, }
exemplifying the need for the 
realizability constraint.

An interesting area for future research would be to devise a method for calculating the optimal 
measurement strategy for Alice and Bob, resulting in the {\blk greatest increase in the purity of the smoothed state relative to that} of the 
filtered state. It would also be useful to investigate \blk how the effectiveness of quantum state smoothing is affected if the putative true state is physically realizable (i.e., corresponds to some unravelling by Bob), 
but the assumed unravelling by Bob was incorrect. Lastly, we have given (Sec.~II) conditions under which quantum smoothing should reduce to classical smoothing, but it is an open question as to whether all of these  conditions are necessary or whether they could be replaced by weaker conditions. \blk 


\begin{acknowledgements}
We acknowledge the traditional owners of the land on which this work was undertaken at Griffith 
University, the Yuggera people. This research is funded by the Australian Research Council Centre of 
Excellence Program CE170100012.  A.C.~acknowledges the support of the Griffith University 
Postdoctoral Fellowship scheme.
\end{acknowledgements}

\appendix
\section{Criteria for smoothed weak-value state and the smoothed quantum state}\label{App-A}
In this appendix we will show how the SWV state and the smoothed quantum state satisfy the properties 
for a quantum state smoothing theory presented in Sec.~\ref{Sec-His}.

\underline{\em Property (1):} {\em \blk The theory should give a single smoothed quantum state $\rho\sm$ 
analogous to the classical state $\wp\sm$, and not a pair of states, for example.}
It is obvious that the SWV state and the smoothed quantum state both satisfy this criterion by their 
respective definitions 
\begin{align}
&\varrho\swv = \frac{\hat{E}\rfil\circ\rho\fil}{\Tr[\hat{E}\rfil\circ\rho\fil]}\,, \\
&\rho\sm = \sum_{\past{\bf{U}}}\wp(\past{\bf{U}}|\both{\bf O})\rho_{\past{\bf O}\past{\bf{U}}}\,.
\end{align}

\underline{\em Property (2):} {\em \blk The smoothed state $\rho\sm \equiv \rho_{\both{\bf O}}$ should 
reduce to its corresponding filtered state after averaging over all possible future (observed) measurement 
records given a past measurement record.} We will first consider the 
SWV state. Averaging the SWV state over the future measurement record given the past record gives
\begin{align}
\sum_{\fut{\bf O}} \wp(\fut{\bf O}|\past{\bf O}) \varrho\sm & = \sum_{\fut{\bf O}} \wp(\fut{\bf O}|\past{\bf O}) 
\frac{\hat{E}\rfil\circ\rho\fil}{\Tr[\hat{E}\rfil\circ\rho\fil]}\\
& = \sum_{\fut{\bf O}}\hat{E}\rfil\circ\rho\fil\,,
\end{align}
where we have used the fact that 
$\Tr[\hat{E}\rfil\circ\rho\fil] \equiv \wp(\fut{\bf O}|\rho\fil) = \wp(\fut{\bf O}|\past{\bf O})$.
By expanding the Jordan product and the completeness relationship for the effect, 
\beq
\sum_{\fut{\bf O}} \hat{E}\rfil = \hat{1}\,,
\eeq
where $\hat{1}$ is the identity operator, we obtain 
\beq
\sum_{\fut{\bf O}} \wp(\fut{\bf O}|\past{\bf O}) \varrho\sm = \rho\fil\,.
\eeq

Now for the smoothed quantum state, averaging over the future observed record conditioned on the past 
observed record gives
\begin{align}
\sum_{\fut{\bf O}} \wp(\fut{\bf O}|\past{\bf O}) &\rho\sm = \sum_{\fut{\bf O}} \wp(\fut{\bf O}|\past{\bf O})
\sum_{\past{\bf{U}}}\wp(\past{\bf{U}}|\both{\bf O})\rho_{\past{\bf O}\past{\bf{U}}}\\
&= \sum_{\past{\bf{U}}}\sum_{\fut{\bf O}} \wp(\fut{\bf O}|\past{\bf O})
\wp(\past{\bf U}|\past{\bf O},\fut{\bf O})\rho_{\past{\bf O}\past{\bf{U}}}\\
&= \sum_{\past{\bf{U}}}\sum_{\fut{\bf O}} \wp(\fut{\bf O}|\past{\bf O})
\frac{\wp(\past{\bf U},\fut{\bf O}|\past{\bf O})}{\wp(\fut{\bf O}|\past{\bf O})}\rho_{\past{\bf O}\past{\bf{U}}}\\
& = \sum_{\past{\bf{U}}} \wp(\past{\bf U}|\past{\bf O})\rho_{\past{\bf O}\past{\bf{U}}} \equiv \rho\fil\,,
\end{align}
where we have used the definition of $\rho\sm$ and $\rho\fil$ in \erf{cond_state} and Bayes' theorem.

\underline{\em Property (3):} {\em \blk The smoothed quantum state should reduce to its classical 
counterpart when the initial conditions, final conditions, and dynamics of the system can all be described 
probabilistically in a fixed basis.}
We assume that the filtered state and the retrofiltered 
effect are diagonal in a fixed orthonormal basis $\ket{\psi_x}$, which we can represent them as 
$\rho\fil = \sum_x \wp(x|\past{\bf O}) \ket{\psi_x}\bra{\psi_x}$ and 
$\hat{E}\rfil = \sum_{x'} \wp(\fut{\bf O}|x')\ket{\psi_{x'}}\bra{\psi_{x'}}$
respectively. To calculate the SWV state, assuming that the system is Markovian, i.e.,~satisfying 
$\wp(\fut{\bf O}|x') \equiv \wp(\fut{\bf O}|x',\past{\bf O})$, we can first compute
\begin{align}
\rho\fil\hat{E}\rfil &= \sum_{x,x'}\wp(x|\past{\bf O})\wp(\fut{\bf O}|x')\ket{\psi_x}
\langle\psi_x|\psi_{x'}\rangle\bra{\psi_{x'}} \\
&= \sum_{x,x'} \wp(x|\past{\bf O})\wp(\fut{\bf O}|x',\past{\bf O}) \delta_{xx'}\ket{\psi_x}\bra{\psi_{x'}}\\
&= \sum_x \wp(x|\both{\bf O})\wp(\fut{\bf O}|\past{\bf O})\ket{\psi_x}\bra{\psi_x}\,.
\end{align}
It is easy to see that the reverse ordering $\hat{E}\rfil\rho\fil$ gives the same result, and we can  
calculate the SWV state
\begin{align}
\varrho\swv &= \frac{\sum_x \wp(x|\both{\bf O})\wp(\fut{\bf O}|\past{\bf O})\ket{\psi_x}\bra{\psi_x}}
{\Tr[\hat{E}\rfil\circ\rho\fil]}\\
&= \sum_x \wp(x|\both{\bf O}) \ket{\psi_x}\bra{\psi_x}\,,
\end{align}
where $\Tr[\hat{E}\rfil\circ\rho\fil] = \wp(\fut{\bf O}|\past{\bf O})$.
We can see that the SWV state has reduced to the classical smoothed state 
$\wp\sm(x) = \wp(x|\both{\bf O})$.

We will now show that the smoothed quantum state also reduces to the classical smoothed state when the 
{\em true} state is diagonal in a fixed orthonormal basis. Firstly, we note that we can represent the true 
state by
\beq
\rho_{\past{\bf O}\past{\bf U}} = \sum_x \wp(x|\past{\bf O},\past{\bf U}) \ket{\psi_x}\bra{\psi_x}\,.
\eeq
We also notice that 
$\wp(x|\past{\bf O},\past{\bf U}) = \wp(x|\both{\bf O},\past{\bf U})$, since the true state contains the 
maximum information about the system, and conditioning the system on anymore information cannot 
influence the state. By taking this into consideration, we can write the smoothed quantum state as 
\begin{align}
\rho\sm &= \sum_{\past{\bf U}} \wp(\past{\bf U}|\both{\bf O}) \sum_x \wp(x|\both{\bf O},\past{\bf U}) 
\ket{\psi_x}\bra{\psi_x}\\
&= \sum_x \sum_{\past{\bf U}} \wp(\past{\bf U}|\both{\bf O}) \wp(x|\both{\bf O},\past{\bf U}) \ket{\psi_x}
\bra{\psi_x}\\
&= \sum_x \wp(x|\both{\bf O}) \ket{\psi_x}\bra{\psi_x}\,.
\end{align}

\underline{\em Property (4):}{\em \blk The smoothed quantum state must alway be Hermitian and positive 
semidefinite, that is, it must be a $\mathfrak{S}$-class quantum state.}
Beginning with the SWV state, we can see that this operator is Hermitian by taking the Hermitian conjugate
\beq
\varrho\sm\dg = \frac{\hat{E}\rfil\circ\rho\fil}{\Tr[\hat{E}\rfil\circ\rho\fil]} = \varrho\sm\,,
\eeq
since $\hat{E}\rfil\dg = \hat{E}\rfil$. However, the SWV state cannot be guaranteed to be 
positive semidefinite. 
The product of two positive semidefinite operators $\rho\fil$ and $\hat{E}\rfil$ is only guaranteed to be 
positive semidefinite if they commute, which is not necessarily the case.
For the smoothed quantum state, it is clear that the state is both Hermitian and positive semidefinite as it is 
a mixture of true quantum states, which must be Hermitian and positive semidefinite. 

{\blk
\section{Proof that \erf{FiltCond} is sufficient to generate an \Sclass~smoothed state} \label{App-B}
For the smoothed state to be an $\mathfrak{S}$-class state, it must satisfy: 
$V\sm +i\hbar\Sigma/2 \geq 0$. Considering a putative true covariance $\tilde{V}\god$ and the 
smoothed covariance in \erf{VS}, we would like to show that
\beq \label{SVs}
[(V\fil - \tilde{V}\god)\inv + (V\rfil + \tilde{V}\god)\inv]\inv + \tilde{V}\god +\frac{i\hbar}{2} \Sigma \geq 0\,,
\eeq
is true for any \Sclass~$\tilde{V}\god$ that satisfies \erf{FiltCond}. Since $\tilde{V}\god$ is an \Sclass~state, 
it must satisfy $\tilde{V}\god + i\hbar\Sigma/2 \geq 0$ and as a result, for $V\sm$ to be an \Sclass~state, 
we only require that 
\beq\label{red-prob}
[(V\fil - \tilde{V}\god)\inv + (V\rfil + \tilde{V}\god)\inv]\inv \geq 0\,.
\eeq
It is always the case that $(V\rfil + \tilde{V}\god)\inv \geq 0$ since $V\rfil$ and $\tilde{V}\god$ are 
individually positive semidefinite and, by assumption, it is that case that $V\fil - \tilde{V}\god \geq 0$. 
Consequently, provided the necessary inverses exist, \erf{red-prob} is true for any putative true covariance 
that satisfies \erf{FiltCond}. Thus the smoothed state be an \Sclass~quantum state.

\section{Proof that the smoothed covariance always fits {\blk within} the filtered covariance}\label{App-C}
We want to show that $V\fil - V\sm \geq 0$ for any putative true state $\tilde{V}\god$. To begin, consider
\begin{align}
V\fil - V\sm =& V\fil - [(V\fil - \tilde{V}\god)\inv + (V\rfil + \tilde{V}\god)\inv]\inv - V\god\\
=& (V\fil - \tilde{V}\god) - (V\fil - \tilde{V}\god)\times\nn\\
&[I + (V\rfil + \tilde{V}\god)\inv(V\fil - \tilde{V}\god)]\inv\\
=& (V\fil - \tilde{V}\god) - (V\fil - \tilde{V}\god) \times\nn\\
&\{I - [I + (V\rfil + \tilde{V}\god)\inv(V\fil - \tilde{V}\god)]\inv\times\nn\\
&(V\rfil + \tilde{V}\god)\inv(V\fil - \tilde{V}\god)\}\label{identity}\\
=& (V\fil - \tilde{V}\god)[I + (V\rfil + \tilde{V}\god)\inv(V\fil - \tilde{V}\god)]\inv \times\nn\\
&(V\rfil + \tilde{V}\god)\inv (V\fil - \tilde{V}\god)\,,
\end{align}
where in \erf{identity} we have used the identity $(I+P)\inv = I - (I+P)\inv P$.
Since $(V\fil - \tilde{V}\god)$ is symmetric, $V\fil - V\sm$ will be positive semidefinite if
\beq
[I + (V\rfil + \tilde{V}\god)\inv (V\fil - \tilde{V}\god)]\inv(V\rfil + \tilde{V}\god)\inv\geq 0\,.
\eeq
\blk Now, \blk
\begin{align}
&[I + (V\rfil + \tilde{V}\god)\inv (V\fil - \tilde{V}\god)]\inv(V\rfil + \tilde{V}\god)\inv \\& = 
[(V\rfil + \tilde{V}\god)\{I + (V\rfil + \tilde{V}\god)\inv(V\fil - \tilde{V}\god)\}]\inv\\ 
& = [V\rfil + V\god + V\fil - V\god]\inv\\
& = [V\rfil + V\fil]\inv \geq 0\,,
\end{align}
where the last line follows from the fact that $V\fil \geq 0$ and $V\rfil \geq 0$. Hence, we have proven that 
for any putative true state $\tilde{V}\god$ that  $V\fil - V\sm \geq 0$. From this we can also show, trivially, 
that $V^{\rm ss} - V\sm \geq 0$ since $V^{\rm ss} - V\fil \geq 0$, i.e.,
\beq
V^{\rm ss} - V\sm = (V^{\rm ss} - V\fil) + (V\fil - V\sm) \geq 0\,.
\eeq
}


\end{document}